# A compact and reconfigurable silicon nitride time-bin entanglement circuit

C. Xiong,[1,*] X. Zhang,[1] A. Mahendra,[1,2] J. He,[1] D.-Y. Choi,[3] C. J. Chae,[4] D. Marpaung,[1] A. Leinse,[5] R. G. Heideman,[5] M. Hoekman,[5] C. G. H. Roeloffzen,[6] R. M. Oldenbeuving,[6] P. W. L. van Dijk,[6] C. Taddei,[7] P. H. W. Leong,[2] and B. J. Eggleton,[1]

[1]*Centre for Ultrahigh bandwidth Devices for Optical Systems (CUDOS), Institute of Photonics and Optical Science (IPOS), School of Physics, University of Sydney, NSW 2006, Australia*

[2]*School of Electrical and Information Engineering, University of Sydney, NSW 2006, Australia*

[3]*CUDOS, Laser Physics Centre, Australian National University, Canberra, ACT 2601, Australia*

[4]*Advanced Photonics Research Institute, Gwangju Institute of Science and Technology, Gwangju, Korea*

[5]*LioniX B.V., P.O. Box 456, 7500 AL, Enschede, The Netherlands*

[6]*SATRAX B.V., HTF, Veldmaat 10, 7522 NM, Enschede, The Netherlands*

[7]*Laser Physics and Nonlinear Optics group, MESA+ Institute for Nanotechnology, University Twente, P.O. Box 217, 7500 AE, Enschede.*

*Corresponding author: chunle@physics.usyd.edu.au*

**Photonic chip based time-bin entanglement has attracted significant attention because of its potential for quantum communication and computation. Useful time-bin entanglement systems must be able to generate, manipulate and analyze entangled photons on a photonic chip for stable, scalable and reconfigurable operation. Here we report the first time-bin entanglement photonic chip that integrates time-bin generation, wavelength demultiplexing and entanglement analysis. A two-photon interference fringe with an 88.4% visibility is measured (without subtracting any noise), indicating the high performance of the chip. Our approach, based on a silicon nitride photonic circuit, which combines the low-loss characteristic of silica and tight integration features of silicon, paves the way for scalable real-world quantum information processors.**

Entanglement is at the heart of photonic quantum technologies such as secure communication [1], super-resolution metrology [2], and powerful computation [3]. Photons are usually entangled in one of three degrees of freedom: polarization, optical path, or time bin. On-chip polarization entangled photon sources have been reported [4, 5], but only the components for photon generation were on-chip due to the difficulty of integrating polarization analysis devices. Chip-scale optical path entangled photon generation and analysis [6], and teleportation [7] have seen rapid development, aiming for on-chip quantum computation. Time-bin entanglement is of particular interest because it (i) can be extended to higher dimensions for computation [8]; (ii) is insensitive to polarization fluctuation and polarization dispersion, and therefore very promising for long distance quantum key distribution (QKD) [1]; and (iii) is naturally compatible with integrated optics: photons can be generated in nonlinear waveguides, entangled and analyzed using on-chip unbalanced Mach-Zehnder interferometers (UMZIs) [9, 10].

For time-bin entanglement to be useful in the real world, the on-chip integration of the entire entanglement system is essential. The high performance of the entanglement system not only relies on photon generation, but also hinges on the compactness, scalability and reconfigurability of the photonic circuit that generates the time bins, demultiplex the photons and analyze the entanglement. Refs [9] and [10] reported photon generation from compact silicon devices, but the wavelength demultiplexing was off chip, and entanglement analysis was based on silica waveguides, which have large bending radii due to their low index contrast. These features are incompatible with high density integration.

In this paper we report, for the first time, a time-bin entanglement photonic chip that integrates time-bin generation, wavelength demultiplexing and entanglement analysis. Our demonstration was based on a high index contrast silicon nitride ($Si_3N_4$) circuit. The waveguide bending radii were reduced from millimeter (for silica) to micrometer scale while maintaining low losses, making high density integration possible. Beyond compactness, the circuit offered tunability and stability, resulting in high performance: a two-photon interference (TPI) fringe with an 88.4% visibility (without subtracting any noise) was measured.

Figure 1(a) shows the principle of time-bin entanglement generation involving four steps: (I) using an UMZI to generate pump 'early' (E) and 'late' (L) time bins with a relative delay of $\delta t$ and phase difference of $\varphi_p$; (II) using a nonlinear waveguide to generate correlated photon pairs, called signal and idler, via a nonlinear process such as spontaneous four-wave mixing (SFWM); (III) using wavelength demultiplexers to remove pump and separate photon pairs; and (IV) using another two UMZIs identical to the first one for entanglement analysis. In step (II) the pump power is controlled so that photon pairs can only be generated either in the 'early' or 'late' time bin, both with a probability of 50%. This forms a superposition state $\frac{1}{\sqrt{2}}(|E\rangle_s|E\rangle_i + e^{j2\varphi_p}|L\rangle_s|L\rangle_i)$, namely time-bin entanglement. In step (IV) when the 'early' photons pass through the longer path and the 'late' photons pass through the shorter, non-classical TPI will occur in the 'middle' (M) time bin because the 'early' and 'late' photons are indistinguishable [8–10].

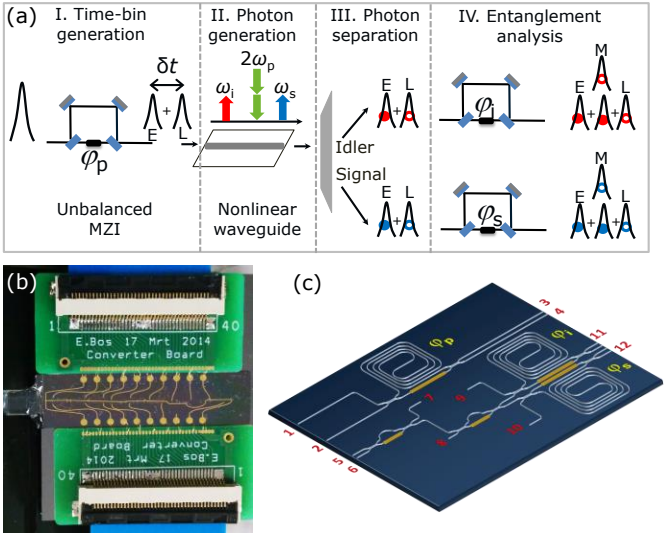

Fig. 1 (a) The principle of time-bin entanglement generation. (b) A photograph of the $Si_3N_4$ based time-bin entanglement chip. The yellow parts are wire bonds for heaters, and the green parts are printed circuit boards for providing voltage to the heaters. On the left hand side is the pigtailed fiber array. (c) The schematic structure of the photonic circuit on the chip. Black lines represent the $Si_3N_4$ waveguides, and the yellow lines are the resistive heaters. Only heaters for pump, signal and idler phase shift, and wavelength demultiplexing are shown. Each tunable coupler consists of a Mach-Zehnder interferometer with a heater in one arm (not shown). In total there are 15 heaters on the chip. The numbers label the ports.

To generate high quality time-bin entangled photons and achieve high visibility interference, three requirements must be met. Firstly, the relative delay between the two time bins must be longer than the minimum of the: pump pulse width, photon coherent time, and resolution of the detection system. This arrangement can avoid single photon interference and ensures that the detection system can distinguish different time bins. On the other hand, a long delay will reduce the bit rate for quantum information processing and can also introduce additional propagation loss. The second requirement is that the relative phase difference $\varphi_{p,s,i}$ between the longer and shorter paths must be stable, and the path length difference of the three UMZIs must be almost equal (errors within the coherent time of laser pulses); otherwise interference will not occur. The third requirement is that the probability for photons appearing in each time bin must be equal to 50% to maximize the interference fringe's visibility. Photonic integration can meet all of these requirements.

In our demonstration, we focused on integrating the components for steps (I), (III) and (IV) on a single chip (28×8 mm) to achieve the abovementioned phase stability, path length accuracy and exact 50% probability of generating photons in each time bin. The circuit was made using the double-stripe $Si_3N_4$ TriPleX™ waveguide technology with LioniX BV [11, 12]. The waveguides consisted of two stripes of $Si_3N_4$ layers stacked on top of each other with $SiO_2$ as an intermediate layer and cladding. The stripes were designed to be 1.2 µm wide, and the $Si_3N_4$ layers and the $SiO_2$ intermediate layer were designed to be 170 and 500 nm thick, respectively. This was optimized for single mode operation at 1550 nm with a high index contrast, allowing a 125 µm bending radius with a propagation loss of <0.2 dB/cm for TE polarization. The photograph and the schematic layout of the chip are shown in Fig. 1(b) and (c). The longer arm of the unbalanced MZIs was approximately 14 cm long and made in a spiral fashion, benefiting from the small bending radius offered by $Si_3N_4$. This gave 795 ps delay relative to the shorter. The demultiplexers consisted of two Mach-Zehnder interferometers (MZIs): the first rejected the pump by varying phase $\varphi_{f1}$ and the second separated signal and idler photons by adjusting phase $\varphi_{f2}$.

All MZIs and UMZIs incorporated tunable couplers for input and output. Each tunable coupler is a balanced MZI having two directional couplers with a fixed ratio close to 50:50 and a phase shifter in one arm to achieve arbitrarily tunable splitting ratio. Tunability was critical for generating and analyzing photons in two time bins with equal probabilities, and for achieving lossless demultiplexing. All phases were controlled through resistive heaters that were wire-bonded (yellow lines in Fig. 1b) to standard electronic printed circuit boards. Independent 16-bit digital-to-analog converters provided high-resolution control of 15 heaters on the chip. When a voltage ($U$) is applied, the temperature of the waveguide under the heater will exponentially increase to a maximum value determined by the power dissipated in the heater. Accordingly, the refractive index of the waveguide will increase and this will introduce phase shift. This thermal-optic phase shift changes quadratically with $U$ [7]. All waveguides, except the longer arm of the UMZIs, were a few mm long so that the heaters can be constructed on top to achieve any phase shift up to $2\pi$ while maintaing low loss and keeping within the dissipation limits of the heaters. All optical input and output ports of the chip were arranged to align with a waveguide array with a spacing of 127 µm and pigtailed to a polarization maintaining fiber (PMF) array (not shown in Fig. 1c). A straight reference waveguide was included for alignment and coupling loss measurement during the pigtailing process. The total insertion loss of the 6.65 cm long reference waveguide was measured to be 4.5 dB.

Prior to the quantum entanglement experiment, we characterized the circuit in the classical regime to ensure that the couplers and demultiplexers were set correctly. For the UMZI that was used to generate two pump time bins (on the top of Fig. 1c, labelled by $\varphi_p$), we injected a pulse train with a pulse width of 10 ps at 1555.7 nm to the input port 2 and monitored them from the two output ports 3 and 4 using a fast oscilloscope. Because of the delay in the longer arm, we observed double pulses separated by 795 ps at each output. The early pulses from both outputs were from the shorter arm and the late ones were from the longer arm. By adjusting the voltage applied to the heater for the output coupler, the amplitudes of the early pulses became equal, as did the late pulses when the output splitting ratio was exactly 50:50. This was independent of the input splitting ratio. As the pulses from the longer arm experienced higher loss, the input splitting ratio deviates from 50:50. By adjusting the voltage applied to the heater for the input coupler, the early pulses became equal to the late pulses in amplitude when the effective input splitting ratio was 50:50 after taking into account the longer arm loss.

For demultiplexer 1 (labelled by $\varphi_{f1}$) that was used to reject the pump, we could not measure the two outputs because one output was connected to demultiplexer 2 (labelled by $\varphi_{f2}$) on the chip. To characterize it, we injected a broadband amplified spontaneous emission (ASE) source to port 7 and monitored the spectra at ports 5 and 6 using an optical spectral analyzer (OSA). By adjusting the voltages applied to the three heaters for the MZI, the spectra measured from ports 5 and 6 were complementary to each other and showed nearly zero loss at the transmission bands when all heaters were set properly. The pink trace in Fig. 2 shows the spectrum taken at port 5 when the heaters for demultiplexer 1 were optimized. The black, blue and red traces indicate the signal, pump and idler wavelengths used in our experiments. The isolation to the pump is at least 25 dB. Due to symmetry of a MZI, when we inject pump, signal and idler to port 6 in the quantum experiment, the input from demultiplexer 1 to 2 will keep signal and idler photons and reject the pump.

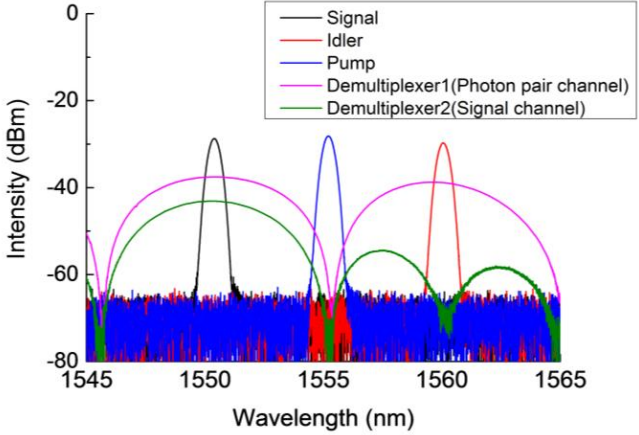

Fig. 2 Characterization of the wavelength demultiplexers. The black, blue and red traces show the signal, pump and idler wavelengths. The pink trace shows the rejection of pump at the photon pair channel from demultiplexer 1, and the green trace shows the signal channel from demultiplexer 2.

Because both outputs of demultiplexer 2 are connected to UMZIs on-chip, we characterized them (labelled $\varphi_i$ and $\varphi_s$) using the approach for setting up the pump UMZI before we could optimize demultiplexer 2. For characterizing demultiplexer 2 we injected an ASE source to port 6 and monitored the spectra from ports 11 and 12. When all heaters for demultiplexer 2 were set correctly, the spectra from ports 11 and 12 showed a dip at the pump wavelength and were complementary at the signal and idler wavelengths. The green trace in Fig. 2 shows the spectrum taken from port 12, indicating very good separation of signal and idler channels. Because the output couplers for the two UMZIs were 50:50, the green trace shows a 3 dB loss in the transmission window at the signal wavelength. The total loss at the transmission windows of the whole circuit was measured to be approximately 10 dB, which comprised fiber-waveguide coupling loss, propagation loss in the longer arm and a 3 dB loss in the UMZI output coupler. It should be noted from Fig. 2 that the MZI based demultiplexers cannot do narrow-band filtering, but can separate pump, signal and idler very well. This is sufficient for performing quantum operations on signal and idler photons on-chip with off-chip narrow-band filters placed before single-photon detectors (SPDs). Once the SPDs are on chip, narrow-band filtering on-chip is required and this is possible with the TriPlex technology [12].

After all heaters were set in place, we performed the time-bin entanglement experiments. The setup is illustrated in Fig. 3. The pump was a mode-locked fiber laser emitting 10 ps pulses at 1555.7 nm with a repetition rate of 50 MHz. The pulses were injected into the first UMZI from port 2 for pump time-bin generation. The output from port 3 of the UMZI was coupled to a 3 mm long, 220 nm high and 460 nm wide silicon nanowire on another chip for photon pair generation. The output of the nanowire was sent back to port 6 of the on-chip demultiplexers, which rejected the pump and separated the signal (1550.9 nm) and idler (1560 nm) photons. Both outputs of demultiplexer 2 were connected to the UMZIs on-chip for entanglement analysis.

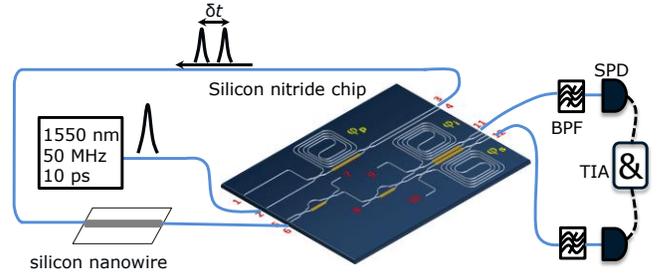

Fig. 3 Experimental setup. Blue solid lines are optical fibers, and black dashed lines are electric cables. BPF: band-pass filter, TIA: time-interval analyzer, and SPD: single-photon detector.

The signal and idler photons were coupled to off-chip band-pass filters (PBFs) through ports 12 and 11 to further remove the pump and be post-selected in the 0.5 nm bandwidth shown in Fig. 2, before being detected by two InGaAs avalanche SPDs ( ID210 from ID Quantique). The SPDs were gated by the 50 MHz laser clock and the gates were aligned with the 'middle' time bin. The effective gate width was 1 ns. We used this SPD configuration to avoid the detection of photons from other time bins [9]. To minimize the dark counts and after-pulsing probability, the detection efficiency was set at 10% and deadtime was set at 20 μs. The coincidences were analyzed by a time interval analyzer (TIA).

The coincidences were a function of $\cos(\varphi_s+\varphi_i+2\varphi_p)$ [9, 10]. At the coupled peak power of 0.45 W (into the silicon nanowire), when we fixed $\varphi_p$ and $\varphi_i$, and varied $\varphi_s$ by adjusting the voltage ($U$) applied to the heater, we observed an 86.8% visibility fringe, without subtracting any noise (Fig. 4, dark squares). To confirm the entanglement, high-visibility fringes must be observed in two non-orthogonal measurement bases [9, 10]. We thus slightly changed $\varphi_i$ by applying 4 V voltage to its heater and again varied $\varphi_s$ which resulted in an 88.4% visibility fringe (red dots) being observed. Both fringes exhibited high visibilities beyond the classical limit of 70.7%, clearly indicating that high quality time-bin entanglement had been achieved.

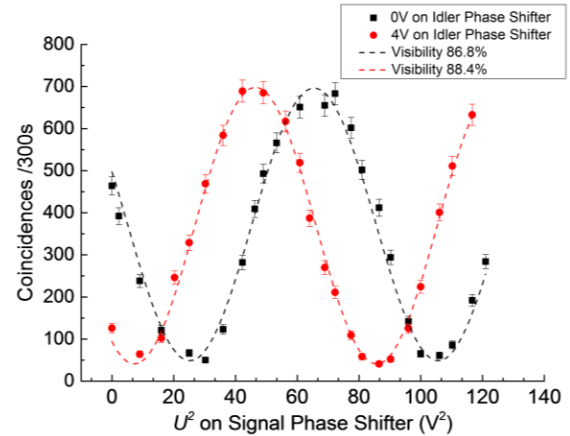

Fig. 4 Coincidences as a function of the square of the applied voltage to the heater of the UMZI in the signal photon channel. Black squares and red dots represent two non-orthogonal measurements when the voltage applied to the heater of the UMZI in the idler channel was set at 0 and 4 V, respectively. The dashed lines are cosine fits. Poisson error bars are used.

While recording the coincidence measurements, we monitored the singles count at each detector and observed that they were insensitive to $\varphi_{s,i}$ (Fig. 5), further evidence that the interference fringe was due to two-photon entanglement.

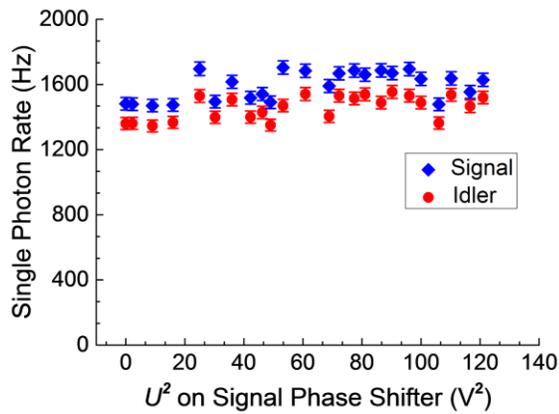

Fig. 5 The measured singles for the signal (blue diamonds) and idler (red dots) channels at different applied voltages to the signal channel heater. Poisson error bars are used.

The visibility of an ideal TPI fringe should be 100%. Our demonstration shows a maximum 88.4% visibility because of accidental coincidences from a few noise sources. The first is the dark count of the detectors. The second is due to the small probability of producing multiple pairs. Finally, the third is due to the so-called charge persistence effect of the SPDs [13, 14]. This effect means when SPDs work at the gated mode, the photons that arrive at the SPDs will interact with the detector even though the gate is OFF. The interaction will trap electrons and produce a detection pulse once the gate is back ON in a few nanoseconds. In our measurements, we observed an extra small coincidence peak besides the main peak in the histogram, indicating the capture of photons from the early time bin even though the detectors were expected to only detect the photons from the middle time bin.

Although our demonstration did not integrate the nonlinear waveguide for photon pair generation on the same chip, recent progress has shown that it is possible to use $Si_3N_4$ nonlinear devices for photon pair generation [15], and such nonlinear devices can be made using the TriPlex technology [16]. In addition, a $Si_3N_4$ platform compatible with our circuit exhibits fast stress-optic effect at >1 MHz [17], indicating the potential of our circuit operating at much higher speed in the future for quantum information processing. Our demonstration clearly establishes that $Si_3N_4$ photonic circuits incorporating all four steps illustrated in Fig. 1(a) for time-bin entanglement are feasible. Once this technology is available, taking advantage of the reconfigurability of the circuit, on-chip time-bin entangled photons can be tested locally as demonstrated in this paper, and then be switched to port 7 for long distance entanglement distribution by simply controlling the phase $\varphi_{f1}$. On the other hand, taking advantage of the compactness of the circuit, multiple time-bin entanglement sources and Bell measurement devices can be made onto a single chip for on-chip time-bin qubits teleportation.

In conclusion we have demonstrated a high performance time-bin entanglement photonic chip based on $Si_3N_4$. This is a significant step towards the ultimate goal of completely integrating all components to realize chip-scale time-bin qubit transmitters and receivers for QKD, and integrating many entanglement sources and analysis circuits on a chip for large-scale quantum computation.

**ACKNOWLEDGMENT**

This work was supported by the Australian Research Council (ARC) Centre of Excellence (CUDOS, CE110001018), Laureate Fellowship (L120100029), Future Fellowship (FT110100853), and the Discovery Early Career Researcher Award programs (DE120100226 and DE150101535). We would like to thank Alex Clark for providing financial support to the chip fabrication through DE130101148.